\documentclass[showpacs,amsmath,amssymb,onecolumn,prx,superscriptaddress,nofootinbib]{revtex4-1}

\usepackage[dvips]{graphicx} 
\usepackage{amsfonts}
\usepackage{amssymb}
\usepackage{amscd}
\usepackage{amsmath}    
\usepackage{enumerate}
\usepackage{epsfig}
\usepackage{subfigure}
\usepackage{xcolor}
\usepackage{amsthm}
\usepackage{slashbox}

\begin{document}

\title{A simple implementation of quantum key distribution based on single-photon Bell state measurement}

\author{Wen-Ye Liang}
\author{Mo Li}
\author{Zhen-Qiang Yin}
\email{yinzheqi@mail.ustc.edu.cn}
\author{Wei Chen}
\email{kooky@mail.ustc.edu.cn}
\author{Shuang Wang}
\author{Xue-Bi An}
\author{Guang-Can Guo}
\author{Zheng-Fu Han}
\affiliation{Key Laboratory of Quantum Information, University of Science and Technology of China, Hefei 230026, China\\
and Synergetic Innovation Center of Quantum Information $\&$ Quantum Physics, University of Science and Technology of China,\\
Hefei, Anhui 230026, China}

\begin{abstract}
Recently some alternatives of the measurement device independent quantum key distribution(MDI-QKD) based on the single-photon Bell state measurement (SBSM) have been proposed. Although these alternatives are not precisely as secure as  MDI-QKD, they possess the advantage of high key rate of traditional BB84-like protocol and avoid the technical complexity of two-photon interference required in the MDI-QKD. However, the setups of these proposed schemes are rather complicated compared to commonly used BB84 systems. Here we propose a simple implementation of SBSM-based QKD which is directly built on the existing realization of BB84 QKD. Our proposal exhibits the hidden connection between SBSM-based QKD and traditional phase-coding QKD protocols. This finding discloses the physics behind these two different types of QKD protocols. In addition, we experimentally demonstrate the feasibility of our protocol.
\end{abstract}

\maketitle


\textbf{Introduction.}
 Quantum key distribution(QKD) is the first quantum technology that comes into real life. It is aimed at sharing unconditionally secure information between two communication parties Alice and Bob even in the presence of a malicious eavesdropper Eve, which is impossible for classical strategy. Since its first proposal in 1984\cite{BB84}, the theoretical analysis and practical implementation of QKD has been universally and deeply studied\cite{security1, security2, qkd1, qkd2, qkd3, qkd4, F-M, decoyexperiment1, decoyexperiment3, decoyexperiment4, decoyexperiment5, decoyexperiment6, decoyexperiment7, F-M}. Recently some significant improvements on the security of QKD's practical implementation are proposed. For example, all the detector related loop-holes can be removed by the measurement device independent QKD (MDI-QKD) protocol\cite{MDI1, MDI2}. MDI-QKD protocol has attracted worldwide attention and has been widely studied both in theory \cite{dtheory1,dtheory2,dtheory3,dtheory4,dtheory5,MDIMCS} and experimental implementations \cite{MDIEXP1,MDIEXP2,MDIEXP3}. And even loop-holes in the state preparation stage for MDI-QKD can be removed in qubit case\cite{dtheory4}.

\par However, an implementation of MDI-QKD protocol requires the interference of photons from two individual lasers, which is still relatively complicated for present technology and leads to a substantial experimental decrease in the key rate in comparison with the BB84 protocol. For alleviating this side effect, there are some alternative schemes proposed recently\cite{SMDIQKD1,SMDIQKD2,SMDIQKD3}. Instead of utilizing the two-photon interference, these schemes utilize a single-photon Bell state measurement setup which avoids the difficulty of two-photon interference and possess the property of high key production of the traditional QKD scheme. In these schemes (we call them SBSM-QKD for abbreviation, since they are based on single-photon Bell state measurement), Alice first encodes a photon in a two-dimension space, polarization for example. Then Alice sends the photon to Bob who further encodes the photon in another degree of freedom, such as spatial path and time bin. In the end, Bob measures the incoming photon with an untrusted Bell state measurement setup.
\par In all previous implementations of SBSM-QKD, the polarization encoding and path encoding are both conducted. Thus its implementation is more complicated than a BB84's. In this article a simple realization of the idea of SBSM-QKD based on the phase encoding QKD setup is given. In our scheme, only time-bin and path encoding are needed. Hence it is very concise and quite similar as conventional phase coding BB84 systems. At last, by considering that only one Bell state projection will suffice the security of MDI-QKD or SBSM-QKD, it is found that a SBSM-QKD can be even realized with a setup which is totally identical to that of phase encoding BB84 system without any modification.

Before proceeding to our proposal, one must note that the fact that as a simpler and more practical alternative to the previous proposals, SBSM-QKD is not strictly as secure as MDI-QKD, because assumptions on measurement devices must be made in SBSM-QKD \cite{Qi's comment}. However, SBSM-QKD may be more secure in some sense, compared to traditional QKD protocols, like BB84. For example, the security of SBSM-QKD can be proved even under the assumption that Eve can decide the output of the BSM device \cite{SMDIQKD1}. Thus, as a simple way to implement SBSM-QKD, our proposal is also not a MDI-secure one. The merit of our scheme is that one may achieve it even using the existing phase-coding QKD systems. In the next section, we give our scheme and prove our scheme is equivalent to the one proposed in Ref.\cite{SMDIQKD2} mathematically.

\textbf{A simple implementation.}
In a SBSM-QKD, the most essential part is the measurement of four Bell states:
\begin{equation}
\label{bellstate}
\begin{split}
|\Phi\rangle^{\pm}={1\over\sqrt{2}}(|00\rangle\pm|11\rangle) \\
|\Psi\rangle^{\pm}={1\over\sqrt{2}}(|01\rangle\pm|10\rangle)
\end{split}
\end{equation}
or any other states combination of equivalent mathematical form.

\begin{figure}[htbp]
\centering\includegraphics[width=8cm]{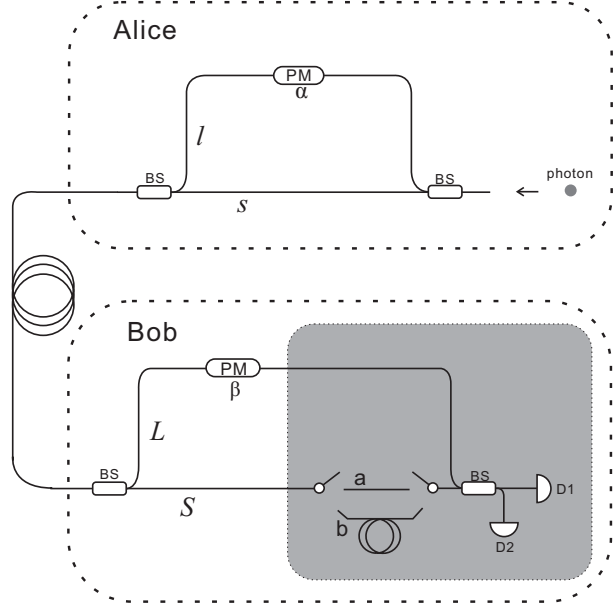}
\caption{\label{Schematic}Schematic of the modified phase encoding protocol for SBSM-QKD. BS: beamsplitter(fiber). PM: phase modulator. SPD: single photon detector. The difference between our protocol and the original phase encoding is the optical switches in Bob's short arm which are designed to select different paths at different timing window. a detection event at the time $|lS\rangle$ or $|sL\rangle$ indicates the projection onto $|\Psi\rangle^{\pm}={1\over\sqrt{2}}(|lL{\rangle}\pm|sS{\rangle})$ and a detection event at the time $|lL\rangle$ indicates the projection onto $|\Phi\rangle^{\pm}={1\over\sqrt{2}}(|lS{\rangle}\pm|sL{\rangle})$. The '+' or '-' depends on which detector signals.}
\end{figure}

Here we use the phase encoding QKD setup as shown in Fig.1 to show how Alice and Bob to encode and conduct the Bell state measurement(BSM). For simplicity, we assume Alice is equipped with an ideal single photon source. In the figure, the shaded area is untrusted area while the other is trusted and characterized. To realize a SBSM-QKD, Alice first chooses the encoding phase $\alpha$ randomly from the set \{$0, \pi/2, \pi, 3\pi/2$\} and apply it to her single photon. The phase is introduced only on the long arm $l$ while no modulator is placed on the short arm $s$. This is essentially an encoding on the spatial basis of Alice. The photon state leaving Alice's region has the form:
\begin{equation}
|\phi\rangle={1\over\sqrt{2}}(|l{\rangle}e^{i\alpha}+|s{\rangle}).
\end{equation}
In Bob's area, the photon is further encoded with his spatial basis, that is, path $L$ and $S$. Similar to Alice's operation, an additional phase $\beta$ chosen from \{$0, \pi/2, \pi, 3\pi/2$\} is introduced in the long arm $L$. Since Alice and Bob encode independently, it is trivial to write the photon state after encoding operations:
\begin{equation}
\label{hybrid}
|\phi\rangle={1\over\sqrt{2}}(|l{\rangle}e^{i\alpha}+|s{\rangle}){\otimes}{1\over\sqrt{2}}(|L{\rangle}e^{i\beta}+|S{\rangle}).
\end{equation}
Obviously, the encoding operations performed by Alice and bob are totally equivalent to the encoding process in Ref.\cite{SMDIQKD2}. One should also note that this encoding process is just the same as the experiment made in Ref.\cite{decoyexperiment7}.

Next the photon is sent to the BSM device for measurement, which is the shaded area in the figure.
We will show the BSM in our scheme is also equivalent to Ref\cite{SMDIQKD2}'s.
\par The Eq.(\ref{hybrid}) can be expanded into the following:
\begin{equation}
\label{expanded}
|\phi\rangle={1\over\sqrt{2}}({1\over\sqrt{2}}(|lL{\rangle}e^{i(\alpha+\beta)}+|sS{\rangle})+{1\over\sqrt{2}}(|lS{\rangle}e^{i\alpha}+|sL{\rangle}e^{i\beta})).
\end{equation}
With different combination of $\alpha$ and $\beta$ chosen from the set \{$0, \pi/2, \pi, 3\pi/2$\}, it can be seen that in Eq.(\ref{expanded}) the former half in the bracket, $(|lL{\rangle}e^{i(\alpha+\beta)}+|sS{\rangle})/\sqrt{2}$, is virtually the $|\Phi\rangle^{\pm}$ in Eq.(\ref{bellstate}) and the latter half, $(|lS{\rangle}e^{i\alpha}+|sL{\rangle}e^{i\beta})/\sqrt{2}$, is the $|\Psi\rangle^{\pm}$ in Eq.(\ref{bellstate}). In the original phase encoding protocol, only $|\Psi\rangle^{\pm}$ is measured for generating keys, which is virtually a partial BSM. To obtain a complete BSM, it is necessary to measure all of the four Bell states. For this purpose, a path selection structure is added to the short arm in Bob's area to delay the $|sS\rangle$, as shown in the figure. The length of path $a$ guarantees that $|sL\rangle$ and $|lS\rangle$ interfere at the beam-splitter in BSM device and can be measured at time $t0$ by single detectors, while the length of $b$ guarantees that $|lL\rangle$ and the delayed $|sS\rangle$ also interfere at the beam-splitter in BSM device and can be measured at a later time $t1$.
\begin{table}[tbp]
\caption{The detection events with respect to different phase application by Alice and Bob, and the corresponding Bell states. The photon traveling $|lS\rangle$ or $|sL\rangle$ arrive at $t0$ while the photon traveling $|lL\rangle$ or the delayed $|sS\rangle$ arrives at $t1$.}
\centering
\begin{tabular}{|c|c|c|c|c|}
 									    \multicolumn{5}{c}{detection at $t0$:} 		       \\  \hline						
 Alice/Bob &         0 &                          $\pi$ & 			$\pi/2$ & 			$3\pi/2$   \\ \hline
 0            &	    D1/$\Psi^{+}$	&	D2/$\Psi^{-}$		&	-		     &			-	\\   \hline
 $\pi$ 	&	    D2/$\Psi^{-}$	&	D1/$\Psi^{+}$	&	-		     &			-	\\   \hline
 $\pi/2$    &		-			&		-			&	D1/$\Psi^{+}$&	D2/$\Psi^{-}$      \\   \hline
 $3\pi/2$	 &		-			&		-			&	D2/$\Psi^{-}$ &	D1/$\Psi^{+}$     \\   \hline
 \end{tabular}
 \begin{tabular}{|c|c|c|c|c|}
 									    \multicolumn{5}{c}{detection at $t1$:} 				\\  \hline		
 Alice/Bob &         0 &                          $\pi$ & 			$\pi/2$ & 			$3\pi/2$    \\  \hline
 0            &	          D1/$\Phi^{+}$		&D2/$\Phi^{-}$	   & -                              &-		\\   \hline
 $\pi$ 	&	    	 D2/$\Phi^{-}$            &D1/$\Phi^{+}$       &-                                &-       \\      \hline
 $\pi/2$    &				-			&		-		         &	D2/$\Phi^{-}$&	D1/$\Phi^{+}$ \\    \hline
 $3\pi/2$	 &				-			&		-	                &	D1/$\Phi^{+}$&	D2/$\Phi^{-}$  \\   \hline

 \end{tabular}

\end{table}

\par In summary, our SBSM-QKD protocol works as follows:
\begin{enumerate}
\item Alice encodes the photon with his spatial basis and a phase modulator.
\item Bob encodes the photon with his spatial basis and a phase modulator.
\item The complete Bell measurement is performed. For this purpose, Bob operates the optical switches to select from paths $a$ and $b$ in the following way. At the time $|sS{\rangle}$ arrives, he selects the path $b$. Afterwards, when $|lS{\rangle}$ arrives, he selects the path $a$. After the measurement of $|\Psi\rangle^{\pm}$, he selects again path $b$ to conduct the measurement of $|\Phi\rangle^{\pm}$.
\item Alice and Bob distill secure keys from the detection results where their phases are in the same basis of the original phase encoding protocol, that is, $|\alpha-\beta|=0~or~\pi$.
\end{enumerate}
\par In the protocol, a detection event at the time $t0$ indicates the projection onto $|\Psi\rangle^{\pm}$ and a detection event at the time $t1$ indicates the projection onto $|\Phi\rangle^{\pm}$. The '+' or '-' depends on which detector signals. The specific outcomes are shown in Table I. One can see that this table is totally equivalent to the table I in Ref.\cite{SMDIQKD2} easily. Thus even the BSM device in the shaded area in the Fig.1 is also equivalent to the one in Ref.\cite{SMDIQKD2}. Therefore, our scheme is an implementation of SBSM-QKD protocol.

\textbf{Experimental Demonstration.}
In the following, we experimentally demonstrate the feasibility of SBSM-QKD protocol. Our experiment is mainly based on a Faraday-Michelson phase-encoding QKD system, in which the polarization
disturbances caused by quantum channel and optical devices can be auto-compensated \cite{F-M}. Thus, keeping steady and high interference
fringe visibility is easy in this QKD system.

\begin{figure}
\centering\includegraphics[width=9cm]{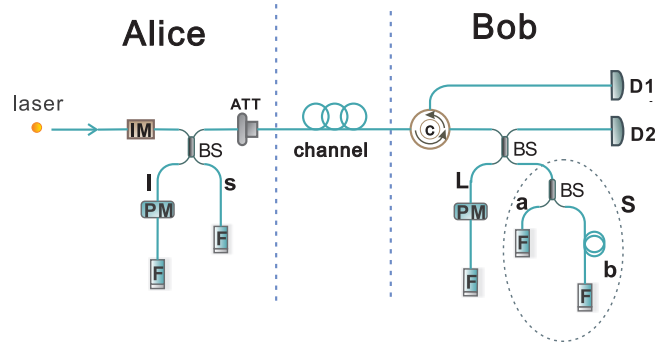}
\caption{(color online). \label{setup}Experimental schematic diagram. BS: beamsplitter(fiber). PM: phase modulator. D1, D2: single photon detector. ATT: attenuator. IM: intensity modulator. In order to simplify the experiment process, we replace the optical switch in Fig. \ref{Schematic}, which is intended for choosing paths of the photon, with a beam splitter. Furthermore, for increasing the stability of the system, we adopt the Faraday-Michelson phase-coding system.}
\end{figure}

\begin{figure}
\centering\includegraphics[width=8cm]{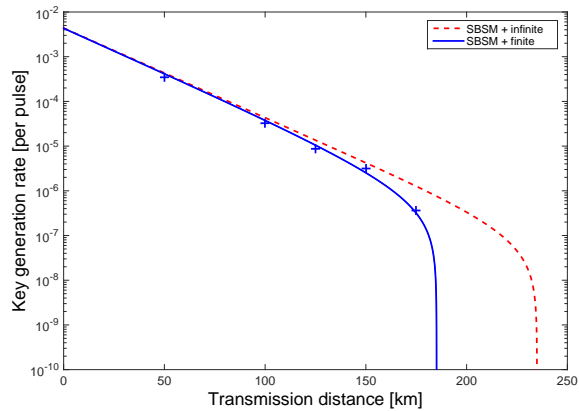}
\caption{(color online). \label{result}Theoretical(lines) and experimental(crosses) key rates in logarithmic scale with respect to the distance. The solid lines correspond to the key rate calculated with $10^{9}$ signal pulses. In comparison, the dotted line represents the key rate with infinite signals. For this two curves we consider the following parameters: security bound $\varepsilon=10^{-8}$, probability that a photon hit the erroneous detector $e_{detector}=0.015$ and correction efficiency $f=1.16$. Other parameters, for instance, the detection efficiency, are the same as experimental parameters. Five crosses represents experimental key rates. At 175 km, the secret key rate is $3.655\times10^{-7}/pulse$.}
\end{figure}

The experiment schematic is shown as Fig. \ref{setup}. In this experiment, we use a laser emitting $1550nm$ weak coherent pulses. The entire setup is synchronically working at a repetition rate of $1MHz$. The intensities of signal, decoy, and vacuum states are $\mu = 0.6$, $\nu = 0.2$, $0$, respectively, and their pulse numbers ratio is 6:2:1. Single-photon detectors in our experiment are Superconducting Nanowire Single Photon Detector (SNSPD), whose dark count probability is approximately $5 \times 10^{-8}/gate$ and the detection efficiency is about $15.3\%$. At each experimental distance we collected $10^{9}$ signal pulses . The loss coefficient of the channel is 0.2 dB/km and the insertion loss is 5 dB on Bob. In addition, secret key rate is estimated according to finite-key analysis proposed in Ref.\cite{MDIFK} which adopted the Chernoff bound. Then we compare the experimental result with theoretically expected result in Fig. \ref{result}. Our results clearly shows that even with a realistic finite size, say $10^{9}$, it remains possible to achieve secure SBSM-QKD at the distances as long as 175km.

\textbf{Conclusion.}
We propose a simple realization of the recently proposed idea of SBSM-QKD, basing on the phase encoding structure while taking a tiny modification. Furthermore, we experimentally demonstrate the feasibility of this protocol. Then with finite-key analysis, we reached a secure distance of 175 km, and the Secret key rate is $3.655\times10^{-7}/pulse$.

A complete Bell state measurement is performed in the protocol, which means that the protocol bears the same mathematical description and security level with the other SBSM-QKD schemes\cite{SMDIQKD1,SMDIQKD2}. In this article, we have a comment on those papers. According to theoretical security analysis on the MDI-QKD, it is known that the measurement of partial Bell state possesses the same security property with the complete Bell state measurement\cite{MDI1,dtheory4,dtheory5}. In Ref.\cite{dtheory2}, it is proved that if the BSM device can only distinguish one Bell state, the security of MDIQKD will remain the same. Thus, it's very reasonable to conjecture that partial BSM will not affect the security of SBSM-QKD protocol. As mentioned in the second section, if we remove the path $b$ of the BSM device in Fig.1, this schematic is just a phase coding BB84 system in Ref.\cite{decoyexperiment7}, which is a partial BSM process.

Hence the original phase encoding, which is essentially a partial Bell state measurement, is equivalently secure to the modified version in this article. Or rather, the original phase encoding scheme as implemented in Ref.\cite{decoyexperiment7}, where Bob modulates four phase-values, is virtually a SBSM-QKD.

\vspace{1\baselineskip}
 The first two authors contributed equally in this letter. The authors thank B. Qi, X.-F. Ma, G. Xavier, M. Curty and G. Lima for helpful discussions. This work was supported by the National Basic Research Program of China (Grants No. 2011CBA00200 and No. 2011CB921200) and National Natural Science Foundation of China (Grants No. 61475148, No. 61201239, and No. 61205118).

\end{document}